\documentclass[12pt]{article}
\usepackage{amssymb}
\begin{document} 


\title{Vacuum fluctuations in a supersymmetric model in FRW spacetime}
\author{
Neven Bili\'c\thanks{Electronic mail:
bilic@thphys.irb.hr}
 \\
Rudjer Bo\v skovi\'c Institute,
POB 180, HR-10002 Zagreb, Croatia 
}

\maketitle

\begin{abstract}
We study a noninteracting supersymmetric model in an expanding FRW spacetime.
A soft supersymmetry breaking induces a nonzero contribution to the vacuum energy density.
A  short distance cutoff of the order of Planck length
provides a scale for the vacuum energy density comparable with the observed cosmological constant.
Assuming the presence of a dark energy substance in addition to the vacuum fluctuations of the field
an effective equation of state is derived in a selfconsistent approach.
The effective equation of state is sensitive to the choice of the cut-off
but no fine tuning is needed.
\end{abstract}



\section{Introduction}

It is generally accepted that the cosmological constant term
which was introduced ad-hoc in the Einstein-Hilbert action
is actually related to the vacuum energy density of matter fields.
Observational evidence for an
  accelerating expansion \cite{perlmutter,bennett,spergel},
implies that
 the vacuum energy density
dominates
the total energy density today.
The vacuum energy density estimated in a simple quantum field theory
is by about 120 orders of magnitude larger than the value required by
astrophysical and cosmological observations \cite{weinberg1} so that 
extreme fine tuning is needed in order to make a cancellation up to 
120 decimal places.
Theoretically, it is possible that the cosmological constant is precisely zero
and the acceleration of the universe expansion is attributed to
the so called {\em dark energy} (DE), a fluid with sufficiently negative pressure,
such that its magnitude exceeds 1/3 of the energy density.
 Nevertheless, even if such a substance exists, it is extremely difficult
to tune the  vacuum energy to be exactly zero.
Hence,  the fine tuning problem perisits 
unless there exists a symmetry principle that forbids a nonzero vacuum energy.
Such principle is indeed provided by supersymmetry \cite{wess}.
In field theory with exact supersymmetry  the
contributions  of fermions and bosons to  vacuum energy precisely
cancel \cite{weinberg2}. However, the supersymmetry in real world  is not exact.

A nonzero cosmological constant implies the de Sitter 
symmetry group of spacetime rather than the Poincar\'e group which is the spacetime symmetry group of
an exact supersymmetry.
Hence,
the structure of de Sitter spacetime
automatically breaks the supersymmetry.
Conversely, a low energy supersymmetry breaking
could in principle generate a  nonzero cosmological constant of an
acceptable magnitude.
Unfortunately, the scale of supersymmetry breaking required by the particle physics phenomenology
must be of the order of 1 TeV or larger implying a cosmological constant too large
by about 60 orders of magnitude.
Some non-supersymmetric models with equal number of boson and fermion degrees of
 freedom have been constructed \cite{kamenshchik} so that
 all the divergent contributions to the vacuum energy density cancel and a small finite contribution
can be made comparable with the observed value
of the cosmological constant.

In this paper we investigate the fate of vacuum energy
when an unbroken supersymmetric model 
is embedded in spatially flat, homogeneous and isotropic spacetime.
In addition, we assume the presence of a dark energy type of substance obeying the equation of state
$p_{\rm DE}=w\rho_{\rm DE}$, with $w<0$.
Unlike in flat spacetime, the vacuum energy density turns out to be nonzero depending 
on background metric. Hence, the expansion is caused by a combined effect of both
DE and vacuum fluctuations of the supersymmetric field.
Solving the Friedman equations selfconsistently we
find the effective equation of state of DE.
In particular, we find the conditions for which the effective expansion becomes of de Sitter type.
The contribution of the supersymmetric field fluctuations
is found to be of the same order of magnitude as DE and no fine tuning is needed.

We do not claim that our model describes a realistic scenario
but it is tempting to speculate along the lines described in an earlier paper
\cite{bilic} where a naive model of supersymmetry in de Sitter spacetime has been considered.
Our working assumption is that the universe today contains DE and no matter apart from fluctuations of
a supersymmetric vacuum as a relict of symmetry breaking in the early universe.
Since the global geometry is non flat,
the lack of Poincare symmetry will lift the Fermi-Bose degeneracy
and the energy density of vacuum fluctuations will
be nonzero.
 This type of ``soft'' supersymmetry breaking is similar to the supersymmetry breaking at finite temperature 
where the Fermi-Bose degeneracy is lifted by quantum statistics (\cite{Kratzert:2003cr} and references therin).

The remainder of the paper is organized as follows. 
In section \ref{model} we introduce a supersymmetric model in an expanding FRW universe.
The calculations and results are presented in section \ref{calculations}.
In section \ref{effective} we discuss the effective DE equation of state.
Concluding remarks are given 
in section \ref{conclude}.
In appendix \ref{covariant} we review  the covariant regularization schemes
of the vacuum expectation value of the energy momentum tensor
in flat spacetime.

\section{The model}
\label{model}
Here we consider a noninteracting Wess-Zumino supersymmetric model with $N$ species and calculate the energy density of vacuum fluctuations
in de Sitter spacetime.
 In general, the supersymmetric Lagrangian $\cal L$ 
for $N$ chiral superfields has the form \cite{bailin}
\begin{equation}
{\cal L} =  \sum_i \Phi_i^\dag \Phi_i |_D +W(\Phi)|_F + \rm h.c.\, ,
\label{eq001}
\end{equation}
where the index $i$ distinguishes the various left chiral superfields $\Phi_i$
and $W(\Phi)$ denotes the superpotential for which we take
\begin{equation}
W(\Phi) = \frac{1}{2}  \sum_i m_i \Phi_i\Phi_i \, .
\label{eq002}
\end{equation}
Eliminating auxiliary fields by equations of motion
the Lagrangian (\ref{eq001}) may be recast in the form
\begin{equation}
{\cal L} =  \sum_i\left(\partial_\mu \phi_i^\dag \partial^\mu\phi_i -m_i^2 |\phi_i|^2
+\frac{i}{2}\bar{\Psi}_i\gamma^\mu \partial_\mu\Psi_i -\frac{1}{2} m_i\bar{\Psi}_i\Psi_i\right) ,
\label{eq003}
\end{equation}
where $\phi_i$ are the complex scalar and $\Psi_i$ the Majorana spinor fields.
For simplicity, from now on we suppress the dependence on the species index $i$.

Next we assume a curved background spacetime geometry with metric $g_{\mu\nu}$.
Spinors in curved spacetime are conveniently treated using the
so called vierbein formalism.
The metric is decomposed as
\begin{equation}
g_{\mu\nu}(x) =\eta_{ab} {e^a}_\mu {e^b}_\nu ;
\hspace{1cm}
g^{\mu\nu}(x) = \eta^{ab} {e_a}^\mu {e_b}^\nu ,
\label{eq101}
\end{equation}
where the set of coefficients ${e^a}_\mu$ is  called the {\em vierbein}
and
\begin{equation}
{e_a}^\mu=\eta_{ab}g^{\mu\nu}{e^b}_\nu 
\label{eq107}
\end{equation}
is the inverse of the vierbein.
Obviously,
\begin{equation}
g\equiv \det g_{\mu\nu} = -(\det {e^a}_\mu )^2 .
\label{eq201}
\end{equation}
The action  
 may be written as 
\begin{equation}
S= \int d^4x \sqrt{-g}( {\cal L}_B + {\cal L}_F) ,
\label{eq004}
\end{equation}
where ${\cal L}_B$ and ${\cal L}_F$ are the boson and fermion Lagrangians, respectively.
The Lagrangian for a complex scalar field may be expressed as the sum of the
Lagrangians for two real fields 
\begin{equation}
{\cal L}_B =  \frac{1}{2}\sum_{i=1}^2 \left( g^{\mu\nu}  \varphi^i_{,\mu} \varphi^i_{,\nu} 
-m^2 \varphi^{i\,2}\right).
\label{eq005}
\end{equation}
The fermion part is given by \cite{birrell}
\begin{equation}
{\cal L}_F = 
\frac{i}{4}\left(\bar{\Psi}\tilde{\gamma}^\mu\Psi_{;\mu}-
\bar{\Psi}_{;\mu}\tilde{\gamma}^\mu \Psi\right)-
\frac{1}{2} m\bar{\Psi}\Psi ,
\label{eq006}
\end{equation}
where $\tilde{\gamma}^\mu$ are the curved spacetime gamma matrices 
\begin{equation}
\tilde{\gamma}^\mu = {e_a}^\mu \gamma^a ,
\label{eq007}
\end{equation}
with  ordinary Dirac gamma matrices denoted by $\gamma^a$.
Variation of (\ref{eq004}) with respect to $\bar{\Psi}$ yields the Dirac equation
in curved spacetime
\begin{equation}
i\tilde{\gamma}^\mu\Psi_{;\mu}
- m\Psi =0.
\label{eq106}
\end{equation}
The covariant derivatives of the spinor are defined as
\begin{equation}
\Psi_{;\mu}=\Psi_{,\mu} - \Gamma_\mu \Psi ,
\label{eq008}
\end{equation}
\begin{equation}
\bar{\Psi}_{;\mu}=\bar{\Psi}_{,\mu} +\bar{\Psi} \Gamma_\mu \, ,
\label{eq009}
\end{equation}
where 
\begin{equation}
\Gamma_\mu= \frac{1}{8}{\omega_\mu}^{ab}[\gamma^a,\gamma^b] \, ,
\label{eq010}
\end{equation}
with the spin connection \cite{parker}
\begin{equation}
{\omega_\mu}^{ab}=-\eta^{bc}{e_c}^\nu ({e^a}_{\nu,\mu}-\Gamma^\lambda_{\mu\nu}{e^a}_\lambda) .
\label{eq011}
\end{equation}
In FRW metric the vierbein is diagonal and in spatially flat FRW spacetime takes a simple form
\begin{equation}
 {e^a}_\mu = {\rm diag} (1,a,a,a).
\label{eq111}
\end{equation}
where $a=a(t)$ is the cosmological expansion scale.

\section{Calculation of the vacuum energy density and pressure}
\label{calculations}
A spatially flat FRW metric is given by
\begin{equation}
ds^2= dt^2 - a(t)^2 d\vec{x}\,^2 .
\label{eq000}
\end{equation}
It is convenient to  work in the conformal frame with metric
\begin{equation}
ds^2= a(\eta)^2(d\eta^2 - d\vec{x}\,^2),
\label{eq012}
\end{equation}
where the  proper time $t$ of the isotropic observers, or cosmic time, is related to the conformal time $\eta$ as
\begin{equation}
dt= a(\eta) d\eta.
\label{eq013}
\end{equation}

In order to calculate the energy density and pressure of the vacuum fluctuations 
we need the vacuum expectation value of the energy-momentum tensor.
The energy-momentum tensor is 
 derived from $S$ as \cite{birrell}
\begin{equation}
T_{\mu\nu}=
\frac{\eta_{ab}{e^b}_\mu}{\sqrt{-g}}
\frac{\delta S}{\delta {e_a}^\nu} =
T_{\mu\nu}^F+T_{\mu\nu}^B \, ,
\label{eq025}
\end{equation}
where the boson and fermion parts
are derived from the respective scalar and spinor Lagrangians
\begin{equation}
T_{\mu\nu}^B=\sum_{i=1}^2
\partial_\mu\varphi^i\partial_\nu\varphi^i-g_{\mu\nu} {\cal L}_B \, ,
\label{eq125}
\end{equation}
\begin{equation}
T_{\mu\nu}^F=
\frac{i}{4}  \left(\bar{\psi}\tilde{\gamma}_{(\mu}\psi_{;\nu)}-
\bar{\psi}_{(;\mu}\tilde{\gamma}_{\nu)}\psi\right).
\label{eq225}
\end{equation}
Owing to the assumed homogeneity and isotropy of spacetime
the calculation of the density and pressure requires the $T^0_0$ component and the trace
$T^\mu_\mu$.
Specifically
for the  metric  (\ref{eq012}) we obtain
\begin{equation}
{T^B}^0_0={\cal H}_B =  \sum_{i=1}^2\left(\frac{1}{2a^2}(\partial_\eta\varphi^i)^2+ 
\frac{1}{2a^2}(\nabla \varphi^i)^2 +\frac{1}{2}m^2 \varphi^{i\,2}\right),
\label{eq026}
\end{equation}
\begin{equation}
{T^B}^\mu_\mu=  \sum_{i=1}^2\left(-\frac{1}{a^2}(\partial_\eta\varphi^i)^2+ 
\frac{1}{a^2}(\nabla \varphi^i)^2 +2 m^2 \varphi^{i\,2}\right),
\label{eq126}
\end{equation}
\begin{equation}
{T^F}^0_0={\cal H}_F = 
-\frac{i}{4a^4}\left(\bar{\psi}\gamma^j\partial_j\psi-
(\partial_j\bar{\psi})\gamma^j \psi\right)
+\frac{1}{2a^3} m\bar{\psi}\psi ,
\label{eq027}
\end{equation}
\begin{equation}
{T^F}^\mu_\mu= 
\frac{1}{2a^3} m\bar{\psi}\psi .
\label{eq127}
\end{equation}

Assuming a general perfect fluid form of the vacuum expectation value of 
$T_{\mu\nu}$
\begin{equation}
<T_{\mu\nu}>=
(\rho+p)u_\mu u_\nu -p g_{\mu\nu} ,
\label{eq616}
\end{equation}
 the energy density and pressure  of the vacuum fluctuations
 are given by
\begin{equation}
\rho =u^\mu u^\nu <T_{\mu\nu}> ,
\label{eq227a}
\end{equation}
\begin{equation}
p =\frac{1}{3}(\rho\, -<{T^\mu}_\mu>) ,
\label{eq327a}
\end{equation}
where $u_\mu$ is the velocity of the fluid and $< A>$ denotes the vacuum expectation value of
an operator A. 
In particular, for vacuum energy we expect
\begin{equation}
<T_\Lambda^{\mu\nu}>=\rho_\Lambda g^{\mu\nu} ,
\label{eq328}
\end{equation}
in accord with Lorentz invariance.
In this case we have 
\begin{equation}
p_\Lambda=-\rho_\Lambda.
\label{eq330}
\end{equation}
With this equation of state we reproduce empty-space Einstein’s equations with a cosmological constant equal to
\begin{equation}
\Lambda=8\pi G \rho_\Lambda.
\label{eq329}
\end{equation}

In the following sections we make the calculations in comoving coordinates.
 In comoving coordinates 
equations (\ref{eq227a}) and (\ref{eq327a}) simplify to
\begin{equation}
\rho_{\rm vac} =<T^0_0> ,
\label{eq227}
\end{equation}
\begin{equation}
p_{\rm vac} =\frac{1}{3}<T_0^0-T^\mu_\mu>,
\label{eq327}
\end{equation}

\subsection{Scalar fields}
\label{scalar}


Next we consider quantum scalar fields in a spatially flat FRW spacetime
with metric (\ref{eq012}).
Each real  scalar field operator is decomposed as
\begin{equation}
\varphi(\eta , \vec{x}) = \sum_{\vec{k}} a^{-1}\left( \chi_k(\eta)e^{i \vec{k}\vec{x}} a_k
+\chi_k(\eta)^*e^{-i \vec{k}\vec{x}} a_k^\dag \right) ,
\label{eq014}
\end{equation}
in full analogy with the 
standard flat-spacetime expression (\ref{eq514})
 considered in appendix \ref{covariant}.
The function $\chi_k$ satisfies the field equation
\begin{equation}
{\chi}^{\prime\prime}_k+ (m^2a^2+k^2- a''/a) \chi_k=0,
\label{eq015}
\end{equation}
where the prime $'$ denotes a derivative with respect to the conformal time $\eta$.
In massless case the exact solutions to this equation may easily be found \cite{birrell}.
In particular, in de Sitter spacetime $a''/a=1/\eta^2$, and one finds positive frequency solutions 
\begin{equation}
\chi_k =  \frac{1}{\sqrt{2Vk}}e^{-ik\eta}\left(1-\frac{i}{k\eta}\right).
\label{eq016}
\end{equation}
The operators $a_k$ associated with these solutions annihilate the 
adiabatic vacuum  in the asymptotic past (Bunch-Davies vacuum) \cite{birrell,jacobson}.

If $m\neq 0$  solutions to (\ref{eq015}) may be constructed by making use of the 
WKB ansatz 
\begin{equation}
\chi_k(\eta) =  \frac{1}{\sqrt{2VaW_k(\eta)}}e^{-i\int^\eta a W_k(\tau) d\tau} ,
\label{eq017}
\end{equation}
where the function $W_k$ is found by solving (\ref{eq015}) iteratively up to an arbitrary 
order in adiabatic expansion \cite{parker}. For our purpose we need the solution up to
the  2nd order only which reads
\begin{equation}
W_k =  \omega_k +\omega^{(2)} ,
\label{eq117}
\end{equation}
where 
\begin{equation}
\omega_k =  \sqrt{m^2+k^2/a^2} .
\label{eq217}
\end{equation}
The general expression for the second order term 
is \cite{parker}
\begin{equation} 
\omega^{(2)}=
 - \frac{3}{8}\frac{1}{\omega_k}\frac{\dot{a}^2}{a^2}
- \frac{3}{4}\frac{1}{\omega_k}\frac{\ddot{a}}{a}
- \frac{3}{4}\frac{k^2}{a^2\omega_k^3}\frac{\dot{a}^2}{a^2} 
+\frac{1}{4}\frac{k^2}{a^2\omega_k^3}\frac{\ddot{a}}{a}
+\frac{5}{8}\frac{k^4}{a^4\omega_k^5}\frac{\dot{a}^2}{a^2} \, ,
\label{eq317}
\end{equation}
where the overdot denotes a derivative with respect to the cosmic time $t$.
Then, equation (\ref{eq117}) may be written as
\begin{equation}
W_k =  \omega_k -\frac{1}{2\omega_k}\left(
\frac{\dot{a}^2}{a^2}+\frac{\ddot{a}}{a}\right)\left[1+{\cal{O}}(m^2/\omega_k^2)\right] ,
\label{eq417}
\end{equation}
or, using (\ref{eq013}), as
\begin{equation}
W_k =  \omega_k -\frac{1}{\omega_k}
\frac{a''}{a^3}\left[1+{\cal{O}}(m^2/\omega_k^2)\right] .
\label{eq517}
\end{equation}
We can calculate now the vacuum expectation value of the 0-0 component and the trace
of the boson energy-momentum tensor.
Using (\ref{eq128}) and the commutation properties of $a_k$ and $a_k^\dag$,
from (\ref{eq026}) and (\ref{eq126}) with (\ref{eq014}) we find
\begin{equation}
<{T^B}^0_0> =  \frac{V}{a^4}\int \frac{d^3k}{(2\pi)^3}\left( |\chi_k^{\,\prime}|^2+ 
a^2 \omega_k^2| \chi_k|^2\right) ,
\label{eq028}
\end{equation}
\begin{equation}
<{T^B}^\mu_\mu> = -2 \frac{V}{a^4}\int \frac{d^3k}{(2\pi)^3}\left(|\chi_k^{\,\prime}|^2- 
a^2 \omega_k^2|\chi_k|^2 -a^2 m^2|\chi_k|^2\right).
\label{eq228}
\end{equation}
Using(\ref{eq227}) and (\ref{eq017})  with (\ref{eq517}) we obtain
\begin{equation}
\rho^B=\frac{1}{a^3}\int \frac{d^3k}{(2\pi)^3 \omega_k}
\left[\omega_k^2+\frac{1}{2}\frac{a'^{\,2}}{a^4}+
\frac{1}{2}\frac{a'^{\,2}}{a^4}\frac{m^2}{\omega_k^2} +\frac{1}{4}
\left(2\frac{a'^{\,2}a''}{a^7}-\frac{a'a'''}{a^6}\right)\frac{1}{\omega_k^2} 
+{\cal O}(\omega_k^{-4})\right] .
\label{eq030}
\end{equation}
The first term in square brackets is identical to the flat spacetime result. The second term
is a quadratically divergent contribution due to a non flat geometry, the next two terms are 
logarithmically divergent, and the rest is finite.
Similarly, with the help of (\ref{eq327}) we find the boson contribution to the pressure
\begin{eqnarray}
p^B= \frac{1}{a^3}\int \frac{d^3k}{(2\pi)^3\omega_k}
\!\! & \!\!& \!\!
\left[\frac{k^2}{3 a^2}+
\frac{1}{6}\left(3\frac{a'^{\,2}}{a^4}-
2\frac{a''}{a^3}
\right)+ \right. 
\frac{1}{6} \left(3\frac{a'^{\,2}}{a^4}-
\frac{a''}{a^3}
\right)\frac{m^2}{\omega_k^2}
 \nonumber \\
& & 
\left. +\frac{1}{4}\left(2\frac{a'^{\,2}a''}{a^7}-
\frac{a' a'''}{a^6}
\right)\frac{1}{\omega_k^2}
+{\cal O}(\omega_k^{-4}) \right] .
\label{eq130}
\end{eqnarray}

\subsection{Spinor fields}

Next we  proceed to quantize the fermions.   The 
Dirac equation in curved spacetime may be derived from (\ref{eq006}).
 Specifically for a spatially flat FRW metric
we obtain 
 \begin{equation}
i\gamma^0\left(\partial_0+\frac{3}{2}\frac{\dot{a}}{a}\right)\Psi+
i\frac{1}{a}\gamma^j\partial_j\Psi
-m\Psi=0.
\label{eq018}
\end{equation}
Rescaling  the
Majorana fermion field $\Psi$  as
\begin{equation}
\Psi=a^{-3/2}\psi ,
\label{eq019}
\end{equation}
and introducing the conformal time
we obtain for $\psi$   the usual flat spacetime Dirac equation 
\begin{equation}
i\gamma^0\partial_\eta\psi+
i\gamma^j\partial_j\psi
-am\psi=0 ,
\label{eq020}
\end{equation}
with time dependent effective mass $am$.
The quantization of $\psi$ is now straightforward
\cite{baacke,cherkas}.
The Majorana field $\psi$ may be decomposed as usual
\begin{equation}
\psi(\eta , \vec{x}) = \sum_{\vec{k},s} \left( u_{ks}(\eta)e^{i \vec{k}\vec{x}} b_{ks}
+v_{ks}(\eta)e^{-i \vec{k}\vec{x}} b_{ks}^\dag \right),
\label{eq021}
\end{equation}
where the spinor $u_{ks}$ may be expressed as
\begin{equation}
u_{ks}= \frac{1}{\sqrt{V}}\left(
\begin{array}{c}
 (i\zeta_k^{\,\prime}+am\zeta_k ) \phi_s \\
\vec{\sigma} \vec{k} \, \zeta_k \phi_s
\end{array}
\right).
\label{eq022}
\end{equation}
Here, the two-spinors $\phi_s$ are the helicity eigenstates which may be chosen as
\begin{equation}
\phi_+ =\left( \begin{array}{c}
 1 \\
0
\end{array}
\right); \hspace{1cm} 
\phi_- =
\left(\begin{array}{c}
 0 \\
1
\end{array}
\right).
\label{eq023}
\end{equation}
The spinor $v_{ks}$ is related to $u_{ks}$ by charge conjugation 
\begin{equation}
v_{ks}= i \gamma^0\gamma^2 (\bar{u}_{ks})^T .
\label{eq122}
\end{equation}
The norm of the spinors may be easily calculated
\begin{equation}
\bar{u}_{ks}u_{ks}=-\bar{v}_{ks}v_{ks}=
\frac{1}{V}(am \zeta_k^*-i\zeta_k^{*\prime})
(am \zeta_k+i\zeta_k')-\frac{1}{V}k^2|\zeta_k|^2.
\label{eq422}
\end{equation}
The mode functions $\zeta_k$ satisfy the equation
\begin{equation}
\zeta_k''+ (m^2a^2+k^2-im a') \zeta_k=0.
\label{eq024}
\end{equation}
In addition, the functions $\zeta_k$ satisfy the condition
\cite{cherkas}
\begin{equation}
k^2|\zeta_k|^2+ (am \zeta_k^*-i\zeta_k^{*\,\prime})
(am \zeta_k+i\zeta_k^{\,\prime})=C_1 .
\label{eq124}
\end{equation}
It may be easily verified that the left-hand side of this equation is a constant of motion
of equation (\ref{eq024}). 
The constant $C_1$ is fixed by the normalization of the spinors 
and by the  initial conditions. A natural assumption is that
at $t=0$ ($\eta=-1/H$, $a=1$) the solution behaves as a plane wave
$\zeta_k=C_2e^{-iE_kt}$,
where $E_k=\sqrt{k^2+m^2}$.
 This gives $\zeta_k(0)=C_2$, $\zeta_k^{\,\prime}(0)=-iC_2E_k$,
and hence  $C_1=2C_2^2E_k (m+E_k)$. From (\ref{eq422}) and (\ref{eq124}) we obtain 
\begin{equation}
\bar{u}_{ks}u_{ks}=-\bar{v}_{ks}v_{ks}=\frac{1}{V}(C_1
-2k^2|\zeta_k|^2) ,
\label{eq222}
\end{equation}
which at $t=0$ reads  
\begin{equation}
\bar{u}_{ks}u_{ks}=-\bar{v}_{ks}v_{ks}=
C_1\frac{m}{VE_k} \, .
\label{eq322}
\end{equation}
For $ C_1=1$ this coincides with the standard  flat spacetime normalization \cite{birrell}.

In massless case the solutions to (\ref{eq024}) are plane waves.
For $m\neq 0$ two methods have been used to solve (\ref{eq024}) for a general spatially flat FRW spacetime:
a) expanding in negative powers of $E_k$ and solving a recursive set of differential equations
\cite{baacke} b) using  a WKB ansatz similar to  (\ref{eq017}) and the adiabatic expansion \cite{cherkas}.

By making use of the decomposition (\ref{eq021}) and the standard anti-commuting
properties of the creation and annihilation operators, 
the vacuum expectation value of the 0-0 component (\ref{eq027})
and of the trace (\ref{eq127})
of the fermion energy-momentum tensor 
 may be written as
\begin{equation}
 <{T^F}^0_0> = \frac{1}{2a^4}\sum_{\vec{k},s} \bar{v}_{ks}
 (am-\vec{k}\,\vec{\gamma})
v_{ks} \, ,
\label{eq031}
\end{equation}
\begin{equation}
 <{T^F}^\mu_\mu> = \frac{1}{2a^4}\sum_{\vec{k},s}am\bar{v}_{ks}
  v_{ks} \, .
\label{eq131}
\end{equation}
Evaluating the expression under the sum and replacing the sum with an integral as in (\ref{eq128})
we obtain
\begin{equation}
<{T^F}^0_0> =\frac{1}{a^4}\int \frac{d^3k}{(2\pi)^3}\left[ik^2
(\zeta_k\zeta_k^{*\prime} -\zeta_k^*\zeta_k')-am \right] ,
\label{eq032}
\end{equation}
\begin{equation}
<{T^F}^\mu_\mu> =-\frac{1}{a^4}\int \frac{d^3k}{(2\pi)^3}am\left(1
-2k^2|\zeta_k|^2 \right).
\label{eq132}
\end{equation}
The expressions  under the  integral sign in (\ref{eq032}) and (\ref{eq132})
 were calculated by Baacke and Patzold \cite{baacke}.
We quote their results for the divergent contributions:
\begin{equation}
<{T^F}^0_0>_{\rm div}=\frac{1}{a^4}\int \frac{d^3k}{(2\pi)^3}\left[
-E_k -\frac{(a^2-1)m^2}{2E_k}
+\frac{(a^2-1)^2m^4}{8E_k^3}
+\frac{a'^{\,2}m^2}{8E_k^3}
 \right],
\label{eq033}
\end{equation}
\begin{equation}
<{T^F}^\mu_\mu>_{\rm div}=-\frac{1}{a^4}\int \frac{d^3k}{(2\pi)^3}\left[
 \frac{a^2m^2}{E_k}
-\frac{a a''m^2}{4E_k^3}
-\frac{a^4m^4}{2E_k^3}
+\frac{a^2m^4}{2E_k^3}
 \right].
\label{eq133}
\end{equation}
Note that the first three terms in square brackets in (\ref{eq033})
are identical to the first three
terms in the expansion of $a\omega_k=\sqrt{E_k^2 + a^2m^2-m^2}$ in powers of
$E_k^{-2}$.
Hence we can write
\begin{equation}
\rho^F=<{T^F}^0_0>=\frac{1}{a^3}\int \frac{d^3k}{(2\pi)^3\omega_k}\left[
-\omega_k^2 +\frac{1}{8}\frac{a'^{\,2}}{a^4}\frac{m^2}{\omega_k^2}
+{\cal O}(\omega_k^{-4})
 \right].
\label{eq034}
\end{equation}
The first term in square brackets is precisely the  flat spacetime vacuum energy
of the fermion field. 
The second term
is a logarithmically divergent contribution due to the FRW geometry and 
the last term is finite and vanishes in the flat-spacetime limit
$ a' \rightarrow 0$.
Note  that, as opposed to bosons,  there is no quadratic divergence
of the type $a'^{\,2}/\omega_k$.

Similarly, from (\ref{eq133}) we obtain
\begin{equation}
<{T^F}^\mu_\mu>=\frac{1}{a^3}\int \frac{d^3k}{(2\pi)^3\omega_k}\left[
-m^2+\frac{1}{4}\frac{a''}{a^3}\frac{m^2}{\omega_k^2}
+{\cal O}(\omega_k^{-4})
 \right],
\label{eq134}
\end{equation}
and using (\ref{eq327}) we find the fermion contribution to the pressure
\begin{equation}
p^F=\frac{1}{a^3}\int \frac{d^3k}{(2\pi)^3\omega_k}\left[
-\frac{1}{3}\frac{k^2}{a^2}+\frac{1}{24}\frac{{a'}^2}{a^4}\frac{m^2}{\omega_k^2}
-\frac{1}{12}\frac{a''}{a^3}\frac{m^2}{\omega_k^2}
+{\cal O}(\omega_k^{-4})
 \right] .
\label{eq234}
\end{equation}

\subsection{Putting it all together}

Assembling the boson and fermion contributions, 
the final expressions for the vacuum energy density and pressure
of each chiral supermultiplet are
\begin{eqnarray}
\rho=\rho^B+\rho^F
\!\! & \!=\!& \!\!
\frac{1}{a^3}\int \frac{d^3k}{(2\pi)^3\omega_k}
\left[
\frac{1}{2}\frac{a'^{\,2}}{a^4}
+ \right. 
\frac{5}{8} \frac{a'^{\,2}}{a^4}
\frac{m^2}{\omega_k^2}
 \nonumber \\
& & 
\left. +\frac{1}{4}\left(2\frac{a'^{\,2}a''}{a^7}-
\frac{a' a'''}{a^6}
\right)\frac{1}{\omega_k^2}
+{\cal O}(\omega_k^{-4}) \right] ,
\label{eq035}
\end{eqnarray}
\begin{eqnarray}
p=p^B+p^F
\!\! & \!=\!& \!\!
\frac{1}{a^3}\int \frac{d^3k}{(2\pi)^3\omega_k}
\left[
\frac{1}{6}\left(3\frac{a'^{\,2}}{a^4}-
2\frac{a''}{a^3}
\right)+ \right. 
\frac{1}{24} \left(13\frac{a'^{\,2}}{a^4}-
6\frac{a''}{a^3}
\right)\frac{m^2}{\omega_k^2}
 \nonumber \\
& & 
\left. +\frac{1}{4}\left(2\frac{a'^{\,2}a''}{a^7}-
\frac{a' a'''}{a^6}
\right)\frac{1}{\omega_k^2}
+{\cal O}(\omega_k^{-4}) \right] .
\label{eq135}
\end{eqnarray}
The dominant contributions in (\ref{eq035}) and (\ref{eq135}) come from the leading terms in square brackets
which diverge quadratically.
Note that these quadratically divergent terms 
 are due to bosons; 
fermions only provide a cancellation of all divergent and finite terms
in the respective flat spacetime contributions of bosons or fermions.

To make the results finite we need to regularize the integrals.
We will use a simple 3-dim momentum cutoff regularization
(recently dubbed ``brute force'' cut-oof regularization \cite{xue})
which, as shown in appendix \ref{covariant}, may be regarded as a covariant regularization 
in a preferred Lorentz frame defined by the DE fluid.

The advantage of this approach is 
%
%
a clear physical meaning of the regularization scheme:
one discards the part of the momentum integral over those momenta where a different, 
yet unknown physics should occur.
In this scheme a preferred Lorentz frame is invoked which is natural 
in a cosmological context where a preferred reference frame exists: the frame fixed by the CMB background or large scale matter distribution. A similar standpoint was advocated by Maggiore \cite{maggiore}
and Mangano \cite{mangano}.
Furthermore, as we have already demonstrated, a supersymmetry provides
 a  cancellation of all flat spacetime contributions
 irrespective of the regularization method one uses.

We  change the integration variable  
to the physical momentum 
$p=k/a$ and introduce a cutoff of the order of the Planck mass 
$\Lambda_{\rm cut}\sim  m_{\rm Pl}$.
The leading terms yield
\begin{equation}
\rho=\frac{N}{4\pi^2}\frac{a'^{\,2}}{a^4}\int_0^{\Lambda_{\rm cut}}\!\!
p\,d\!p \left(1+{\cal O}(p^{-2})\right)
\cong \frac{N\Lambda_{\rm cut}^2}{8\pi^2}
\frac{a'^{\,2}}{a^4}
\left(1+{\cal O}(\Lambda_{\rm cut}^{-2}\ln{\Lambda_{\rm cut}})\right),
\label{eq036}
\end{equation}
\begin{equation}
p
\cong \frac{N\Lambda_{\rm cut}^2}{24\pi^2}
\left(3\frac{a'^{\,2}}{a^4}-
2\frac{a''}{a^3}
\right)
\left(1+{\cal O}(\Lambda_{\rm cut}^{-2}\ln{\Lambda_{\rm cut}})\right),
\label{eq136}
\end{equation}
where $N$ is the number of chiral species.
Clearly, we do not obtain the vacuum equation of state (\ref{eq330})
as may have been expected as a consequence
of a regularization
that assumes the existence of a preferred Lorentz frame.

In order to estimate the cutoff we first neglect background
DE and assume that
the total energy density $\rho$ is given by (\ref{eq036}).
 If we compare the first  Friedman equation 
with (\ref{eq036}) keeping the leading term on the righthand side 
we find that our  cutoff should satisfy
\begin{equation}
\Lambda_{\rm cut}\cong \sqrt{\frac{3\pi}{N}}\, m_{\rm Pl} \, .
\label{eq037}
\end{equation}
It is worthwhile to note that several  approaches \cite{maggiore,cohen,shapiro,sloth} with substantially different underlying philosophy 
have led to results similar to (\ref{eq036}).
In particular, Cohen, Kaplan, and Nelson
\cite{cohen} have employed
a cosmological horizon radius $R_H=1/H$  as
 a long distance cutoff and
derived an upper bound 
\begin{equation}
\rho\cong \Lambda_{\rm UV}^4\leq \frac{3}{8\pi}\frac{m_{\rm Pl}^2}{L^2} 
 \label{eq038}
\end{equation}
 from  a holographic principle.
Here, $\Lambda_{\rm UV}$ and $L$ denote the ultraviolet and long distance cutoffs, respectively.
Our result would saturate the holographic bound (\ref{eq038})
 if we identify  $a'/a^2=1/L$.

The closest approach to ours is that of Maggiore \cite{maggiore} and Sloth \cite{sloth}
who present a similar calculation
of zero-point energy using massless boson fields only. 
The main difference in \cite{maggiore} with respect to ours is that
the  cancellation of  the quartic contributions
was done by hand on the basis of the procedure used previously in
the literature with the so-called ADM mass. 
In our model, the cancellation of all (not only quartically divergent) 
flat spacetime contributions is naturally provided by supersymmetry.
Another difference is that our results
(\ref{eq036}) and (\ref{eq136}) are sufficiently general to allow
a self consistent approach.

The above consideration gives only an estimate for the cutoff.
In the next section we give a self consistent treatment of the supersymmetric vacuum fluctuations
in the presence of DE.

\section{Effective equation of state}
\label{effective}
Since there is no way to precisely determine the cutoff,
it is convenient to introduce a free dimensionless cutoff parameter $\lambda$
of order $\lambda\lesssim 1$
such that
\begin{equation}
\Lambda_{\rm cut}= \lambda \sqrt{\frac{3\pi}{N}}\, m_{\rm Pl} \, .
\label{eq137}
\end{equation}
The factor $1/\sqrt{N}$ is introduced to make the result independent of
the number of species. 
If we reinstate the cosmic time $t$, equations (\ref{eq036}) and  (\ref{eq136}) become
\begin{equation}
\rho=\lambda \frac{3}{8\pi G}\frac{\dot{a}^2}{a^2}\, ,
\label{eq236}
\end{equation}
\begin{equation}
p=
\lambda \frac{1}{8\pi G}\left(\frac{\dot{a}^2}{a^2}-2\frac{\ddot{a}}{a}\right).
\label{eq336}
\end{equation}
Obviously, the pressure is negative if $\dot{a}^2<2a\ddot{a}$.
E.g., for a de Sitter expansion we find $\dot{a}^2=a\ddot{a}$ and  $p= -\rho/3$.
 This case was considered by Maggiore \cite{maggiore}
who concluded
that the vacuum fluctuations cannot (at least
in his approach) be interpreted as a part of the cosmological constant because
in the second Friedman equation the accelerating effects of pressure are canceled by those from the 
density. We shall see shortly that this conclusion is slightly altered in a selfconsistent 
approach to the effective equation of state.

In addition to vacuum fluctuations of matter fields, we assume existence of DE
characterized by the equation of state $p_{\rm DE}=w\rho_{\rm DE}$.
The Friedman equations then take the form
\begin{equation}
\frac{\dot{a}^2}{a^2}=\frac{8\pi}{3}G\rho_{\rm DE}+\lambda \frac{\dot{a}^2}{a^2} \, ,
\label{eq039}
\end{equation}
\begin{equation}
\frac{\ddot{a}}{a}=-\frac{4\pi}{3}G(\rho_{\rm DE}+3p_{\rm DE}) - \lambda\left( \frac{\dot{a}^2}{a^2}- \frac{\ddot{a}}{a} \right) .
\label{eq040}
\end{equation}
Introducing the effective equation of state
\begin{equation}
p_{\rm eff} = w_{\rm eff} \rho_{\rm eff} \, ,
\label{eq041}
\end{equation}
where
\begin{equation}
\rho_{\rm eff}=\frac{\rho_{\rm DE}}{1-\lambda} \, ,
\label{eq042}
\end{equation}
\begin{equation}
w_{\rm eff}=w+\frac{2}{3}\frac{\lambda}{1-\lambda}\, ,
\label{eq043}
\end{equation}
equations (\ref{eq039}) and (\ref{eq040}) may be recast in the standard FRW form
\begin{equation}
\frac{\dot{a}^2}{a^2}=\frac{8\pi}{3}G\rho_{\rm eff}  \, ,
\label{eq139}
\end{equation}
\begin{equation}
\frac{\ddot{a}}{a}=-\frac{4\pi}{3}G(1+3w_{\rm eff})\rho_{\rm eff} \, .
\label{eq140}
\end{equation}
Three remarks are in order. First, it is clear from (\ref{eq042}) why we have chosen the cutoff parameter $\lambda$  less than 1. Second,  it follows from (\ref{eq043}) that the contribution of the vacuum fluctuations
to the effective equation of state is always positive and hence it goes against acceleration!
The third remark concerns the Bianchi identity which would not be respected if the vacuum fluctuations
were the only source of gravity in Einstein's equations.
However,
because of the additional contribution to the energy-momentum tensor coming from
 DE, it is not necessary to have both contributions separately conserved.
Since the effective pressure and energy density satisfy
Einstein's field equations (\ref{eq139}) and (\ref{eq140}),
the combined energy-momentum is conserved and therfore the Bianchi identity is respected.
In this way an interaction between the  vacuum fluctuations and DE
 is implicitly assumed in the spirit of the two component model of Grande, Sola and 
\v Stefan\v ci\'c. \cite{grande}.

It is worthwhile to analyze interesting  cosmological solutions to
equations (\ref{eq139}) and (\ref{eq140}) depending on
the nature of DE given by the equation of state $p_{\rm DE}=w\rho_{\rm DE}$.
\begin{enumerate}
 \item
Consider first the case when there is no DE, i.e., 
when $p_{\rm DE}=\rho_{\rm DE}=0$.
In this case equations (\ref{eq039}) and (\ref{eq040})
admit only a trivial solution $a=$const.
Clearly, if $\lambda=1$, equation (\ref{eq039}) becomes a trivial identity and
equation   (\ref{eq040}) implies $\dot{a}=0$.
If $\lambda\neq 1$, equations  (\ref{eq039}) and (\ref{eq040}) are satisfied if and only if  $\dot{a}=0$.
Therefore, $a=$const is the only solution to (\ref{eq039}) and (\ref{eq040}) for any choice of $\lambda$.
In other words, FRW spacetime cannot be generated by vacuum fluctuations alone in an empty background.
\item
Another interesting special case is DE represented by a cosmological constant, i.e., 
for the equation of state $p_{\rm DE}= -\rho_{\rm DE}$.
It follows  from (\ref{eq043}) that an accelerated expansion ($w_{\rm eff}<-1/3$) is
achieved for any value of the cutoff parameter in the range  $0 <\lambda <1/2$.
This case has also been discussed in \cite{maggiore,mangano}.
\item
A more general case is obtained  if we only require
accelerating expansion, i.e., if the effective equation of state satisfies
$w_{\rm eff}< -1/3$. 
Then equation (\ref{eq043}) 
implies
that the range $-1 <w< -1/3$  is compatible with
$0<\lambda<1/2$,
 whereas
$w< -1$
  would imply $\lambda>1/2$.
In the latter case the DE equation of state violates 
the dominant energy condition.
The fluid of which the equation of state violates the dominant energy condition
was
dubbed {\em phantom energy} \cite{caldwell1,caldwell2}
and has  recently become a popular alternative
to quintessence and 
cosmological constant \cite{phantom}.
\item
In the last example, we  require that the background be de Sitter , i.e., $w_{\rm eff}=-1$.
In other words 
the effective equation of state describes an effective cosmological constant.
From (\ref{eq043}) we find
\begin{equation}
w= -\frac{2}{3}\frac{\lambda}{1-\lambda} -1 .
\label{eq044}
\end{equation}
Hence, this case may be realized  only for a fluid with $w < -1$, i.e., for the phantom energy. 
We see that in a selfconsistent approach, unlike in the example discussed in \cite{maggiore},
 a de Sitter expansion can be achieved as a result of a combined effect
of DE and vacuum fluctuations.

\end{enumerate}

\section{Conclusion}
\label{conclude}

We have calculated the contribution of supersymmetric fields to vacuum energy 
in spatially flat, homogeneous and isotropic spacetime.
In addition to supersymmetric fields we have assumed existence of a substance obeying the equation of state
$p_{\rm DE}=w\rho_{\rm DE}$, with $w<0$.
Unlike in flat spacetime, the vacuum fluctuations turn out to be nonzero depending 
on background metric. Combining effects of both
dark energy and vacuum fluctuations of the supersymmetric field
in a selfconsistent way we have
found the effective equation of state.
In particular, we have found the conditions for which the effective expansion becomes of de Sitter type.
The contribution of the supersymmetric field fluctuations
is of the same order of magnitude as DE and no fine tuning is needed.

We have found that if we impose a UV cutoff of the order $m_{\rm Pl}$ 
the leading term in the energy density of vacuum fluctuations is of the order
$H^2 m_{\rm Pl}^2$, where $H=\dot{a}/a$.
  In this way, if we identify the  expansion parameter $H$ with the Hubble parameter today,
 the model provides  a phenomenologically acceptable value
of the vacuum energy density.
We have also found that a consistency with the Friedman equations
implies that a natural cutoff must be inversely proportional to $\sqrt{N}$.
A similar natural cutoff  has been recently proposed in order to
resolve the so called species problem of black-hole entropy \cite{dvali}.

\appendix

\section{Covariant regularization of $T_{\mu\nu}$ in flat spacetime}
\label{covariant}
To illustrate problems related to the field theoretical calculation of 
vacuum energy  we review the well known
 results for the scalar field in flat spacetime \cite{ossola,akhmedov,andrianov}.
Consider a single noninteracting real scalar field
 described by the Lagrangian
\begin{equation}
{\cal L} =  \frac{1}{2}  \eta^{\mu\nu}  \varphi_{,\mu} \varphi_{,\nu} 
-\frac{1}{2}m^2\varphi^2 ,
\label{eq500}
\end{equation}
with the corresponding energy-momentum tensor 
\begin{equation}
T_{\mu\nu}=
\partial_\mu\varphi\partial_\nu\varphi-\eta_{\mu\nu} {\cal L} .
\label{eq501}
\end{equation}
The field operator is decomposed as
\begin{equation}
\varphi(t , \vec{x}) = \sum_{\vec{k}} \frac{1}{\sqrt{2VE_k}}\left( e^{-iE_kt +i \vec{k}\vec{x}} a_k
+e^{iE_kt-i \vec{k}\vec{x}} a_k^\dag \right) ,
\label{eq514}
\end{equation}
where 
\begin{equation}
E_k =  \sqrt{m^2+k^2} .
\label{eq515}
\end{equation}
and
$a_k$ and $a_k^\dag$ are the annihilation and creation operators, respectively, 
associated with the plane wave solutions
with the standard commutation properties. 
\begin{equation}
[a_k,a_k^\dag] =\delta_{\vec{k}\vec{k'}}   \, .
\label{eq129}
\end{equation}
From (\ref{eq500})-(\ref{eq514}) with (\ref{eq129})
and replacing the sum over momenta
 by an integral
in the usual way
\begin{equation}
\sum_{\vec{k}} =  V\int \frac{d^3k}{(2\pi)^3} \, ,
\label{eq128}
\end{equation}
we find the vacuum expectation value  of $T_{\mu\nu}$
\begin{equation}
<T_{\mu\nu}>=\frac{1}{2}\int \frac{d^3k}{(2\pi)^3E_k} k_\mu k_\nu \, ,
\label{eq711}
\end{equation}
where $k_\mu=(E_k, \vec{k})$.
The righthand side of (\ref{eq711}) may be expressed in a
manifestly covariant way 
 \cite{ossola}
\begin{equation}
<T_{\mu\nu}>=\int \frac{d^4k}{(2\pi)^3} k_\mu k_\nu \delta (k^\rho k_\rho-m^2)\theta(k_0)   \, .
\label{eq509}
\end{equation}
The delta function under the integral restricts the integration to the hypersurface defined by
\begin{equation}
k^\mu k_\mu-m^2=0  \, ; \hspace{1cm} k_0>0 ,
\label{eq710}
\end{equation}
with the invariant measure $d^3k/E_k$ on the hypersurface.
Performing the integral over $k_0$ in (\ref{eq509})  one 
recovers (\ref{eq711}).
However, if one assumes the vacuum expectation value of $T_{\mu\nu}$ to be of the form
\begin{equation}
<T_{\mu\nu}>=\rho_{\rm vac} g_{\mu\nu} \, ,
\label{eq428}
\end{equation}
as dictated by Lorentz invariance of the vacuum, one  encounters inconsistency since 
different results for $\rho_{\rm vac}$ are obtained depending on  which component of $T_{\mu\nu}$ 
one calculates. 
  For example, using $T_{00}$ one finds
\begin{equation}
\rho_{\rm vac}= <T_{00}>= \frac{1}{2}\int \frac{d^3k}{(2\pi)^3} E_k \, .
\label{eq602}
\end{equation}
 On the other hand, using the trace one finds
\begin{equation}
\rho_{\rm vac}=\frac{1}{4}{T^\mu}_{\mu}= \frac{m^2}{8}\int \frac{d^3k}{(2\pi)^3 E_k}   \, ,
\label{eq604}
\end{equation}
which does not agree with  (\ref{eq602}).
One must conclude that the assumption (\ref{eq428}) is not compatible with (\ref{eq509}).
The reason for this inconsistency is that
the integrals in expressions (\ref{eq711}) and (\ref{eq509}) are divergent 
and   make sense only if they are
regularized.

One way to covariantly regularize (\ref{eq711}) or (\ref{eq509}) is 
to cut the hypersurface (\ref{eq710}) by a spacelike hyperplane defined by
\begin{equation}
f(k_\mu)\equiv u^\mu k_\mu- \sqrt{K^2+m^2}=0 ,
\label{eq601}
\end{equation}
where $K$ is an arbitrary constant of dimension of mass and
$u_\mu$ is
a general future directed timelike unit vector, which may be parameterized as
\begin{equation}
u_\mu=(\cosh \alpha, \sinh \alpha \sin \theta \cos \phi , 
\sinh \alpha \sin \theta \sin \phi\, ,\sinh \alpha \cos \theta) \, . 
\label{eq521}
\end{equation}
Clearly, the vector $u_\mu$ is normal  to the  hypersurface $f=$ const
because $\partial f/\partial k^\mu =u_\mu $.
In this way one effectively introduces a preferred Lorentz frame
defined by the vector $u_\mu$  as if the vacuum fluctuations
are embedded in a homogeneous fluid
moving with the velocity $u_\mu$.
The special form of the constant
 in ({\ref{eq601}) is chosen for convenience.

The hyperplane cuts
the hypersurface (\ref{eq710}) at a 2-dimensional intersection defined by
({\ref{eq601})
together with (\ref{eq710}).
 This gives a quadratic equation
the solutions of which define a 2-dimensional closed surface as a boundary of the integration
domain $\Sigma$ defined by
\begin{equation}
\sqrt{K^2+m^2}- u^\mu k_\mu >0
 \label{eq310}
\end{equation}
together with (\ref{eq710}).
Hence, the regularized expression for $<T_{\mu\nu}> $ is given by
\begin{equation}
<T_{\mu\nu}>=\frac{1}{2}\int_\Sigma \frac{d^3k}{(2\pi)^3E_k} k_\mu k_\nu \, ,
\label{eq701}
\end{equation}
or in a manifestly covariant form
\begin{equation}
<T_{\mu\nu}>=\int \frac{d^4k}{(2\pi)^3}  k_\mu k_\nu \delta (k^\rho k_\rho-m^2)\theta(u^\rho k_\rho)
\theta(\sqrt{K^2+m^2}-u^\rho k_\rho)\, .
\end{equation}
Using a general perfect fluid form (\ref{eq616}), 
$\rho$ and $p$ are given by the invariant expressions
\begin{equation}
\rho =\frac{1}{2}\int_\Sigma \frac{d^3k}{(2\pi)^3 E_k} (u^\mu k_\mu)^2 \, ,
\label{eq717}
\end{equation}
\begin{equation}
 p= \frac{1}{6}\int_\Sigma \frac{d^3k}{(2\pi)^3 E_k} [(u^\mu k_\mu)^2-m^2] \, .
\label{eq718}
\end{equation}
In comoving frame ($\alpha=0$) the integration domain $\Sigma$ becomes a ball of radius $K$
and we obtain
\begin{equation}
\rho = <T_{00}>=\frac{1}{2}\int_{k< K} \frac{d^3k}{(2\pi)^3} E_k \, ,
\label{eq719}
\end{equation}
\begin{equation}
p=<T_{ii}> = \frac{1}{6} \int_{k< K} \frac{d^3k}{(2\pi)^3 E_k} k^2 \, .
\label{eq720}
\end{equation}
Hence, the described covariant regularization is equivalent to a simple
3-dim momentum cut-oof procedure. The integration  yields
\begin{equation}
\rho=\frac{K^4}{16\pi^2}+ \frac{m^2K^2}{16\pi^2}
-\frac{1}{64\pi^2} \ln \frac{K^2}{m^2} + ... \, ,
\label{eq504}
\end{equation}
\begin{equation}
p=\frac{1}{3}\frac{K^4}{16\pi^2}- \frac{1}{3}\frac{m^2K^2}{16\pi^2}
+\frac{1}{64\pi^2} \ln \frac{K^2}{m^2} + ... \, ,
\label{eq505}
\end{equation}
where the ellipses denote the finite terms.

This result reveals two problems.
The first one concerns the fine tuning.
Assuming that the ordinary field theory is valid up to the scale of quantum gravity, 
i.e. the Planck scale, the leading term in (\ref{eq504}) yields 
\begin{equation}
\rho\approx \frac{m_{\rm Pl}^4}{16\pi^2}\approx 10^{73}\, {\rm  GeV}^4 \, ,
\label{eq506}
\end{equation}
compared with the observed value
\begin{equation}
\rho_{\rm cr}\approx  10^{-47}\, {\rm  GeV}^4  \, .
\label{eq507}
\end{equation}
This huge discrepancy may be easily rectified in flat spacetime simply by 
subtracting all divergent contributions and redefining the vacuum
to have its energy exactly zero. 
 However, as soon as we demand that  vacuum energy or cosmological constant is 
 nonzero the calculations should be repeated in curved spacetime (e.g. de Sitter spacetime)
and a simple subtraction of vacuum energy by fiat cannot be done.

If, in addition to the vacuum fluctuations of the field, one assumes that
 there exists an independent cosmological constant term $\Lambda$,
as a result one would find an effective vacuum energy
\begin{equation}
\rho_{\rm eff}=\rho+\rho_{\Lambda} \, .
\label{eq508}
\end{equation}
In order to reproduce the observed value one needs 
a cancellation of the two terms on the right hand side up to 120 decimal places!
The problem is actually much more severe as there are  many
 contributions to vacuum energy from different fields
 with different interactions and all these contribution must 
somehow cancel to give the observed vacuum energy density.

The second problem is related to the equation of state.
Obviously, equations (\ref{eq504}) and (\ref{eq505}) do not reproduce the expected vacuum energy
equation of state (\ref{eq330}), as required by Lorentz invariance.
Instead we find $p=\rho/3$ for the quartic term,  $p=-\rho/3$ for the
quadratic term, and 
only the logarithmic term 
satisfies (\ref{eq330}).
This violation of Lorentz invariance is not surprising
since the adopted covariant regularization procedure assumes 
existence of a preferred Lorentz frame. 

In principle, it is possible to regularize 
the energy momentum tensor by imposing (\ref{eq328}) and 
ignoring the mentioned inconsistency of the derived covariant expression
(\ref{eq711}).
Then,  using  the manifestly covariant form (\ref{eq509}) of  the energy momentum tensor 
one can calculate the components using covariant 
regularization schemes which do not invoke a preferred Lorentz frame.
For example, 
the dimensional regularization 
with the $\overline{\rm MS}$ prescription gives
\cite{akhmedov}
\begin{equation}
\rho_{\rm dim}= -p_{\rm dim} =-\frac{m^4}{64\pi^2}
\left( \ln \frac{K^2}{m^2} + \frac{3}{2}\right) \, ,
\label{eq811}
\end{equation}
and one would  conclude that a covariant regularization removes
the Lorentz violating quartic and quadratic divergences and retains only
the logarithmically divergent term which agrees with the logarithmic
term of 
the 3-dim cutoff procedure
in  (\ref{eq504}) and (\ref{eq505}).
However,
in the Pauli Villars regularization one finds
\cite{ossola}
\begin{equation}
\rho_{\rm PV}= -p_{\rm PV} =\frac{1}{64\pi^2}
\left[-\frac{1}{2} K^4 +2 m^2 K^2 
-m^4\left( \ln \frac{K^2}{m^2} + \frac{3}{2}\right)\right] \, ,
\label{eq510}
\end{equation}
so in this covariant procedure the quartic and quadratic divergences are  present 
with coefficients different from
those of the 3-dim cutoff procedure. Again, the logarithmic term agrees 
with that of (\ref{eq504}) and (\ref{eq505}).
Both dimensional and Pauli Villars regularizations have an unpleasant property that 
the  leading term contribution 
yields $\rho <0$. This property is unphysical since
 $\rho\equiv <T^0_0>$ should be positive for the scalar field as follows from (\ref{eq026}).
Ossola and Sirlin have argued \cite{ossola} that 
 the quartic term in (\ref{eq510}) may be removed by demanding strict scale invariance
in the limit $m\rightarrow 0$ or by invoking the Feynman regulator.

 Two other Lorentz invariant regularization schemes were considered by Andrianov et al \cite{andrianov}:
$\zeta$-function regularization and the UV cutoff regularization of the large wave-number field modes.
It was concluded that the former method is not adequate in treating the cosmological constant problem as it redirects
the problem from UV to IR region.
The latter method with a suitable choice of the large wave-number cutoff reproduces  the Pauli Villars regularization
result (\ref{eq510}). With the choice advocated in \cite{andrianov}
one can get rid off the quartic term but then the coefficient of the quadratic term changes.

We see from the above analysis that a covariant regularization  is ambiguous
although in  all mentioned covariant  methods the logarithmic term 
comes with the same coefficient as in the 3-dim cutoff procedure.
With the exception of the dimensional regularization
where  the power low divergences are absent by definition,
the quadratic term is allways present with a coefficient that depends on the regularization method.

\subsection*{Acknowledgments}
I wish to thank B.\ Guberina, A.Y.\ Kamenshchik, H.\ Nikoli\'c, J.\ Sola, and H.\ \v Stefan\v ci\'c for useful discussions
and comments.
I am particularly indebted to I.\ Shapiro for critical remarks
on the previous version of the paper.
This work was supported by the Ministry of Science,
Education and Sport
of the Republic of Croatia under contracts No. 098-0982930-2864.

\end{document}